\documentclass[aps,amsfonts,amsmath,prd,nofootinbib,tightenlines,superscriptaddress,10pt]{revtex4}

\usepackage{epsfig}
\usepackage{bm}
\usepackage{bbm}
\newcommand{\tr}{\,\mbox{tr}}
\newcommand{\Tr}{\,\mbox{Tr}}
\newcommand{\sech}{\,\mbox{sech}\,}

\newcommand{\sgn}{\,\mbox{sgn}\,}
\begin{document}

\title{Electromagnetic Casimir-Polder Interaction for a Conducting Cone}


\author{Noah Graham}
\email{ngraham@middlebury.edu}
\affiliation{Department of Physics, Middlebury College,
Middlebury, Vermont 05753, USA}

\begin{abstract}
Using the formulation of the electromagnetic Green's function of a
perfectly conducting cone in terms of analytically continued angular
momentum, we compute the Casimir-Polder interaction energy of the cone
with a polarizable particle.  We introduce this formalism by first
reviewing the analogous approach for a perfectly conducting wedge,
and then demonstrate the calculation through numerical evaluation of the
resulting integrals.
\end{abstract}

\maketitle

\section{Introduction}

The Casimir-Polder interaction between an uncharged conducting object
on a polarizable particle
\cite{PhysRev.73.360,PhysRevLett.70.560,PhysRevA.72.033610} provides
one of the simplest examples of a mesoscopic fluctuation-based force.
Since the particle can be treated as a delta-function potential, its
effects can be evaluated in any basis.
As a result, in the scattering formalism, the interaction energy
between the particle and conducting object can be determined directly
from the full electromagnetic Green's function in the presence of the
object.  In contrast, for the Casimir force between two objects, one
needs the scattering $T$-matrices for each object connected by the free
Green's function expressed in each object's scattering basis
to propagate fluctuations between objects
\cite{Langbein1974,Lambrecht06,Emig06,PhysRevLett.97.160401,spheres,PhysRevD.80.085021}.

Along with the standard plane, cylinder, and sphere geometries for
which there exist analytic expressions in terms of scattering modes
for the Green's function in the presence of a perfect conductor, the
conducting wedge \cite{PhysRevD.20.3063,BREVIK1996157}, which
also models a cosmic string \cite{PhysRevD.34.1918}, is a case
where the Green's function can be obtained analytically as a mode sum,
by imposing the wedge boundary conditions through a discrete,
fractional angular momentum index.  However, one can also use analytic
continuation to a continuous, complex angular momentum to express
this Green's function in terms of the $T$-matrix for scattering in the
angular (rather than radial) variable
\cite{Oberhettinger,Carslaw,Felsen,Maghrebi6867}, an approach that
then extends to the case of the cone \cite{Maghrebi6867} and puts the
Green's function into a form that is more directly analogous to the sphere,
cylinder, and plane results. Mathematically, this  approach is based
on the Mehler-Fock and Kontorovich-Lebedev transforms \cite{index_transforms}.
The complex angular momentum approach requires that we consider only
imaginary frequencies, however, so while it is well-suited to
equilibrium problems at both zero and nonzero temperature, it cannot be
applied to heat transfer \cite{PhysRevB.86.115423}, which must be
computed on the real axis.

All of these calculations allow for investigation of the
Casimir-Polder interaction near a sharp edge or tip, where the
derivative expansion approach
\cite{PhysRevD.84.105031,PhysRevD.90.081702,Bimonte_2012,PhysRevA.94.022509,MaterialGradient}
is not applicable, yielding semi-analytic results in terms of a small
number of integrals and sums.  But this approach is limited to
perfect conductors, and as a result complements calculations based on
surface current methods
\cite{Johnson_2007,Reid:2009aa,Johnson_2013,Bimonte_2021},
which are more complex numerically but applicable to more general
geometries and materials.  Recent work using the multiple scattering
surface method, in which one combines expansions in scattering
between and within objects
\cite{PhysRevLett.130.200401} provides a particularly
relevant comparison by demonstrating the Casimir force between a
dielectric wedge and plane.

Here we use the analytically continued scattering formalism to
calculate the Casimir-Polder force of a conducting cone on a
polarizable atom, as might arise, for example, in the case of a
particle beam passing by an atomic force microscope.  We begin by
reviewing the wedge calculation in the discrete angular momentum
approach, and show how to obtain the same result using the analytic
continuation approach.  We then extend this calculation to the case of
the cone, obtaining a result in terms of a sum and integral over
angular momentum variables.  For the special case where the particle
lies on the cone axis, the calculation simplifies to a single
integral.  This calculation can be straightforwardly extended to
frequency-dependent polarizability and nonzero temperature, although
in those cases an additional sum or integral over frequency must be
done numerically.

\section{Review of Casimir-Polder Wedge}

We begin by reviewing the Casimir-Polder interaction energy for a
conducting wedge, which was computed in Refs.\
\cite{PhysRevD.20.3063,BREVIK1998134} and considered in the context of
repulsive forces in Refs.\ \cite{PhysRevA.83.062507,Milton_2012}.  Let the
wedge run parallel to
the $z$-axis and have half-opening angle $0<\theta_0<\pi$ around $\theta=0$
with the wedge vertex located at $x=y=0$,  and consider imaginary
wavenumber $k=i\kappa$ with $\kappa > 0$.  Note that by allowing
$\displaystyle \theta_0 > \frac{\pi}{2}$, we will be able to consider
the case where the particle is inside the wedge.  For a particle
located at angle  $\theta \in [0,2 \pi]$ obeying $\theta_0 < \theta <
2\pi - \theta_0$, one can write the full Green's function for the
wedge in terms of ordinary cylindrical wavefunctions of fractional
order, \cite{PhysRevD.20.3063,BREVIK1996157}
\begin{eqnarray}
\mathfrak{G}(\bm{r}_1,\bm{r}_2,\kappa) &=&
-\frac{p}{\pi} \int_{-\infty}^{\infty} \frac{dk_z}{2\pi}
\sum_{\ell=-\infty}^\infty 
\left(
\bm{M}_{\ell k_z \kappa}^{\rm outgoing} \otimes 
\bm{M}_{\ell k_z \kappa}^{\rm regular}{}^* - 
\bm{N}_{\ell k_z \kappa}^{\rm outgoing} \otimes 
\bm{N}_{\ell k_z \kappa}^{\rm regular}{}^* \right) \,,
\end{eqnarray}
in terms of the magnetic (transverse electric) and electric
(transverse magnetic) modes respectively,
\begin{eqnarray}
\bm{M}_{\ell k_z \kappa}(\bm{r}) &=&
\frac{1}{\sqrt{\kappa^2 + k_z^2}} \nabla \times \left[
\bm{\hat{z}} f_{|\ell p|} \left(\sqrt{\kappa^2 + k_z^2} \, r\right)
e^{i k_z z}  \cos (\ell p(\theta-\theta_0))\right]\cr
\bm{N}_{\ell k_z \kappa}(\bm{r}) &=&
\frac{1}{\kappa \sqrt{\kappa^2 + k_z^2}} \nabla \times \nabla \times
\left[\bm{\hat{z}} f_{|\ell p|} \left(\sqrt{\kappa^2 + k_z^2} \, r\right)
e^{i k_z z} \sin (\ell p(\theta-\theta_0))\right]
\label{eqn:cylmodes}
\end{eqnarray}
with $\displaystyle p=\frac{\pi}{2(\pi-\theta_0)}$, where the regular
(outgoing) function is evaluated at the point $\bm{r}_1$ 
or $\bm{r}_2$ with the smaller (larger) value of the cylindrical
radius $r$ and the radial functions are given in terms of Bessel
functions  for regular and outgoing modes as
\begin{equation}
f^{\rm regular}_{|\ell p|} (\sqrt{\kappa^2 + k_z^2} \, r) = 
I_{|\ell p|} \left(\sqrt{\kappa^2 + k_z^2} \, r\right) 
\qquad \hbox{and} \qquad
f^{\rm outgoing}_{|\ell p|} (\sqrt{\kappa^2 + k_z^2} \, r) =
K_{|\ell p|} \left(\sqrt{\kappa^2 + k_z^2} \, r\right) \,.
\end{equation}
This Green's function then obeys
\begin{equation}
(\nabla \times \nabla \times + \kappa^2)
\mathfrak{G}(\bm{r}_1,\bm{r}_2,\kappa) = \delta^{(3)}(\bm{r}_1-\bm{r}_2)
\end{equation}
in the presence of the conducting wedge, while the free Green's function
$\mathfrak{G}_0(\bm{r}_1,\bm{r}_2,\kappa)$, given by setting
$p=1$ and replacing the trigonometric functions 
$\sin (\ell p(\theta-\theta_0))$ and $\cos (\ell p(\theta-\theta_0))$
in Eq.~(\ref{eqn:cylmodes}) with $\displaystyle \frac{1}{\sqrt{2}}
e^{i \ell \theta}$, obeys the same equation in empty space.

One can then use the ``TGTG''
\cite{Langbein1974,Emig06,Lambrecht06,PhysRevLett.97.160401,spheres,
PhysRevD.80.085021} formulation of the Casimir energy, considering
only the lowest-order interaction with the potential for a particle
with polarizability ${\bm \alpha}$ at position $\bm{r}$,
\begin{equation}
V(\bm{r}') = -4 \pi {\bm \alpha} \kappa^2 \delta^{(3)}(\bm{r}'-\bm{r})\,,
\end{equation}
which can be expressed in any basis since it is a delta-function.
The result for the interaction energy of a particle with isotropic
polarizability $\alpha$ becomes
\cite{PhysRevD.20.3063,BREVIK1998134}
\begin{eqnarray}
U(\bm{r}) 
&=& -\frac{\hbar c}{2\pi} \int_0^\infty \Tr \left[V(\bm{r})
\left(\mathfrak{G}(\bm{r},\bm{r}',\kappa)-\mathfrak{G}_0(\bm{r},\bm{r}',\kappa)
\right) \right]d\kappa \cr
&=& 2\alpha \hbar c \int_0^\infty \kappa^2 \tr \left[
\mathfrak{G}(\bm{r},\bm{r},\kappa)-\mathfrak{G}_0(\bm{r},\bm{r},\kappa)\right]
d\kappa \cr
&=&
-\frac{3\alpha \hbar c}{8 \pi r^4 \sin^4(p (\theta-\theta_0))}
\left[p^4 - \frac{2}{3} p^2 (p^2 - 1) \sin^2(p (\theta-\theta_0)) - 
\frac{1}{135} (p^2 - 1) (p^2 + 11) \sin^4(p (\theta-\theta_0))\right] \,,
\label{eqn:wedge}
\end{eqnarray}
where $\Tr$ includes the trace over the spatial coordinate while $\tr$
is the trace only over polarizations.  In this approach, there is not a
straightforward way to subtract the free contribution mode-by-mode, so
one instead uses a point-splitting argument to subtract the entire
contribution from the free Green's function at once.

For the case of the cone, there does not exist an analog of this full
Green's function, written in terms of a rescaled order.  As a result,
we next recompute the result for the wedge using a different form
of the Green's function, which will generalize more readily to the
case of the cone.  In this approach, the angular momentum sum is
replaced via analytic continuation by an integral, yielding for the
free Green's function \cite{Oberhettinger,Maghrebi6867}
\begin{eqnarray}
\mathfrak{G}_0(\bm{r}_1,\bm{r}_2,\kappa) &=&
-\frac{1}{\pi^2} \int_{-\infty}^{\infty} \frac{dk_z}{2\pi}
\int_0^\infty d\lambda
\left(
\bm{M}_{\lambda k_z \kappa}^{\rm outgoing,+} \otimes 
\bm{M}_{\lambda k_z \kappa}^{\rm regular,+}{}^* +
\bm{M}_{\lambda k_z \kappa}^{\rm outgoing,-} \otimes 
\bm{M}_{\lambda k_z \kappa}^{\rm regular,-}{}^* 
\right. \cr && \left. 
- \bm{N}_{\lambda k_z \kappa}^{\rm outgoing,+} \otimes 
\bm{N}_{\lambda k_z \kappa}^{\rm regular,+}{}^* -
\bm{N}_{\lambda k_z \kappa}^{\rm outgoing,-} \otimes 
\bm{N}_{\lambda k_z \kappa}^{\rm regular,-}{}^* \right) \,,
\end{eqnarray}
where the transverse modes are
\begin{equation}
\bm{M}_{\lambda k_z \kappa}(\bm{r}) = 
\frac{1}{\sqrt{\kappa^2 + k_z^2}} \nabla \times \left[ \bm{\hat z}
K_{i\lambda}\left(\sqrt{\kappa^2 + k_z^2} \, r\right) e^{i k_z z}
f_\lambda(\theta) \right]
\qquad \hbox{and} \qquad
\bm{N}_{\lambda k_z \kappa}(\bm{r}) = \frac{1}{\kappa} \nabla \times
\bm{M}_{\lambda k_z \kappa}(\bm{r})
\end{equation}
and we now take $\theta \in [-\pi,\pi]$.  We have both even and odd modes, 
with regular modes given by
\begin{equation}
f^{\rm regular,+}_\lambda(\theta) = \cosh(\lambda \theta)
\qquad \hbox{and} \qquad
f^{\rm regular,-}_\lambda(\theta) = \sinh(\lambda \theta)
\end{equation}
and outgoing modes given by
\begin{equation}
f^{\rm outgoing,+}_\lambda(\theta) = \cosh(\lambda (\pi - |\theta|))
\qquad \hbox{and} \qquad
f^{\rm outgoing,-}_\lambda(\theta) = \sinh(\lambda (\pi - |\theta|))
\sgn\theta \,,
\end{equation}
where the regular (outgoing) functions are evaluated at the point
$\bm{r}_1$  or $\bm{r}_2$ with the smaller (larger) value of $|\theta|$.
Note that the star indicates conjugation of the complex exponential
part of the function only.  

Although not needed for the computation, the corresponding
longitudinal mode is
\begin{equation}
\bm{L}_{\lambda k_z \kappa}(\bm{r}) =
\frac{1}{\kappa} \nabla \left[
K_{i\lambda}\left(\sqrt{\kappa^2 + k_z^2} \, r\right) e^{i k_z z}
f_\lambda(\theta) \right] \,.
\end{equation}
If its contribution is added to the free Green's function, the result
is equal to the scalar Green's function \newline
$\displaystyle \frac{1}{2 \pi} 
\int_{-\infty}^{\infty} \frac{dk_z}{2\pi}
K_0\left(\sqrt{\kappa^2 + k_z^2} \left|r_1 e^{i\theta_1} - 
r_2 e^{i \theta_2}\right|\right)
e^{ik_z(z_> - z_<)}$ times the identity
matrix; without this contribution, we obtain the same scalar times the
projection matrix onto the transverse components.  Here $z_>$ ($z_<$)
is the $z$ coordinate associated with the point with the larger
(smaller) value of $|\theta|$.

In this approach, we will take the wedge to be located at $\theta
 = \pm \theta_0$ and the particle's location will always have $|\theta| >
 \theta_0$.  We can then obtain the full Green's function by replacing
 the regular solution with a combination of regular and outgoing
 solutions given in terms of the $T$-matrix
\begin{equation}
{\cal T}_{\lambda}^{M,+} = \frac{\sinh (\lambda \theta_0)}
{\sinh (\lambda (\pi-\theta_0))} = - {\cal T}_{\lambda}^{N,-} 
\qquad \hbox{and} \qquad
{\cal T}_{\lambda}^{M,-} = \frac{\cosh (\lambda \theta_0)}
{\cosh (\lambda (\pi-\theta_0))} = - {\cal T}_{\lambda}^{N,+} \,,
\end{equation}
so that it now obeys the conducting boundary conditions on the wedge,
yielding
\begin{eqnarray}
\mathfrak{G}(\bm{r}_1,\bm{r}_2,\kappa) &=&
-\frac{1}{\pi^2} \int_{-\infty}^{\infty} \frac{dk_z}{2\pi}
\int_0^\infty d\lambda
\left[
\bm{M}_{\lambda k_z \kappa}^{\rm outgoing,+} \otimes 
\left(\bm{M}_{\lambda k_z \kappa}^{\rm regular,+}{}^* 
+ {\cal T}_{\lambda}^{M,+} \bm{M}_{\lambda k_z \kappa}^{\rm outgoing,+}{}^*
\right)
\right. \cr && \left. 
+ \bm{M}_{\lambda k_z \kappa}^{\rm outgoing,-} \otimes 
\left(\bm{M}_{\lambda k_z \kappa}^{\rm regular,-}{}^* 
+ {\cal T}_{\lambda}^{M,-} \bm{M}_{\lambda k_z \kappa}^{\rm outgoing,-}{}^*
\right)
\right. \cr && \left. 
- \bm{N}_{\lambda k_z \kappa}^{\rm outgoing,+} \otimes 
\left(\bm{N}_{\lambda k_z \kappa}^{\rm regular,+}{}^* 
+ {\cal T}_{\lambda}^{N,+} \bm{N}_{\lambda k_z \kappa}^{\rm outgoing,+}{}^*
\right)
\right. \cr && \left. 
- \bm{N}_{\lambda k_z \kappa}^{\rm outgoing,-} \otimes 
\left(\bm{N}_{\lambda k_z \kappa}^{\rm regular,-}{}^* 
+ {\cal T}_{\lambda}^{N,-} \bm{N}_{\lambda k_z \kappa}^{\rm outgoing,-}{}^*
\right)
\right] \,.
\end{eqnarray}
In this form we can easily subtract the free Green's function 
mode by mode, leaving only the terms with outgoing waves multiplied by
the $T$-matrix.  We obtain for the energy
\begin{eqnarray}
U(\bm{r}) &=& 
-\frac{ \alpha \hbar c}{\pi ^3 r^2}
\int_0^\infty d\kappa \int_{-\infty}^\infty dk_z
\int_0^\infty d\lambda 
\frac{1}{\sinh(2 (\pi -\theta_0) \lambda) }
\cr &&
\Bigg\{
K_{i \lambda}(\sqrt{\kappa^2+ k_z^2} \, r)^2
    \Bigg[\left(r^2
    \left(\kappa^2 + k_z^2\right)+\lambda ^2\right) 
    \cosh(2(\pi -\theta ) \lambda) \sinh \pi \lambda \cr &&
+\left(r^2
    \left(\kappa ^2+k_z^2\right)+\frac{\kappa^2-k_z^2}{\kappa^2+k_z^2}
    \lambda ^2\right) \sinh((\pi -2 \theta_0)\lambda)\Bigg]
\cr &&
+r^2 \left(\frac{\partial}{\partial r}
K_{i \lambda}(\sqrt{\kappa^2+k_z^2} \, r)
\right)^2
  \left(
\cosh(2 (\pi -\theta) \lambda)
\sinh \pi \lambda + \frac{k_z^2-\kappa^2}{k_z^2+\kappa^2}
  \sinh((\pi - 2 \theta_0) \lambda)  ]\right)
\Bigg\} \cr
&=& -\frac{\alpha \hbar c}{\pi r^4}\int_0^\infty d\lambda
\left[\frac{\lambda + \lambda^3}{3} \coth\pi \lambda - 
\frac{1}{3} \coth (2 (\pi - \theta_0) \lambda) + 
\frac{\cosh (2 (\pi - \theta) \lambda)}
{\sinh (2 (\pi - \theta_0) \lambda)}\right] \,,
\end{eqnarray}
where we have integrated over $\kappa$ and $k_z$ using polar coordinates.
After carrying out the $\lambda$ integral, we obtain agreement with
Eq.~(\ref{eqn:wedge}).

\section{Electromagnetic Cone Green's function}

\begin{figure}[htbp]
\includegraphics[width=0.1875\linewidth]{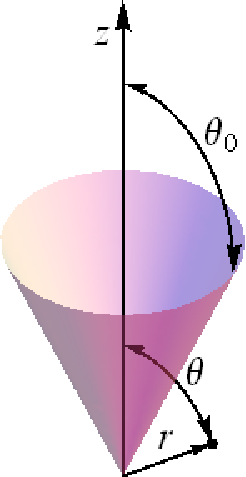}
\caption{Geometry of cone with half-opening angle $\theta_0$ and
particle at radius $r$ and angle $\theta > \theta_0$.}
\label{fig:diagram}
\end{figure}

We now construct the Green's function for the perfectly conducting
cone with half-opening angle $0< \theta_0 < \pi$, centered on the
$z$-axis with the cone vertex at $z=0$, as shown in
Fig.~\ref{fig:diagram}.  We again consider imaginary
wavenumber $k=i\kappa$ with $\kappa > 0$.  Note that by allowing
$\displaystyle \theta_0 > \frac{\pi}{2}$, we can consider the case
where the particle is inside the cone.  From Ref.\ \cite{Maghrebi6867}, we
have magnetic (transverse electric) and electric (transverse magnetic)
transverse modes
\begin{equation}
\bm{M}_{\lambda m \kappa}(\bm{r}) = \nabla \times \left[\bm{r} 
k_{i\lambda-\frac{1}{2}} (\kappa r) e^{im\phi} 
f_{\lambda m} (\bm{r})\right] \qquad \hbox{and} \qquad
\bm{N}_{\lambda m \kappa}(\bm{r}) = \frac{1}{\kappa} 
\nabla \times \bm{M}_{\lambda m \kappa}(\bm{r})
\end{equation}
with
\begin{equation}
f_{\lambda m}^{\rm regular} (\bm{r}) = 
P_{i\lambda-\frac{1}{2}}^{-m}(\cos \theta) 
\qquad \hbox{and} \qquad
f_{\lambda m}^{\rm outgoing} (\bm{r}) = 
P_{i\lambda-\frac{1}{2}}^{m}(-\cos \theta) \,,
\end{equation}
where $P_{i\lambda-\frac{1}{2}}^{m}(\cos \theta)$ is the Legendre
function of the first kind and $\displaystyle
k_{i\lambda-\frac{1}{2}}(\kappa r) = 
\sqrt{\frac{2}{\pi \kappa r}}K_{i\lambda}(\kappa r)$ is the modified
spherical Bessel function of the third kind, both with complex
degree/order $\displaystyle \ell = i\lambda - \frac{1}{2}$.  The
``ghost'' mode \cite{Maghrebi6867} is
\begin{equation}
\bm{R}_{\lambda m \kappa} = \bm{r} \times \nabla 
\left[k_{i\lambda-\frac{1}{2}}(\kappa r) 
P_{i\lambda-\frac{1}{2}}^{-|m|}(\pm \cos \theta) e^{im\phi}\right] \,,
\end{equation}
where the $\pm$ sign is for regular and outgoing modes respectively.
Its contribution arises from the contour integral used to
turn the sum over the angular momentum quantum number $\ell$ into the
integral over its analytic continuation $\lambda$, in which it cancels
the contribution from the $\ell=0$ mode since that mode does not exist
in electromagnetism.  As a result, it is only ever evaluated at
$\displaystyle \lambda = \frac{1}{2i}$, corresponding to $\ell=0$.

In this basis, the free Green's function is \cite{Maghrebi6867}
\begin{eqnarray}
\mathfrak{G}_0(\bm{r}_1,\bm{r}_2,\kappa) &=&
-\frac{\kappa}{4\pi} \Bigg[
\sum_{m=-\infty}^\infty 
\int_0^\infty \frac{\lambda \tanh \pi \lambda}
{\lambda^2 + \frac{1}{4}} d\lambda \left(
\bm{M}_{\lambda m \kappa}^{\rm outgoing} \otimes 
\bm{M}_{\lambda m \kappa}^{\rm regular}{}^* - 
\bm{N}_{\lambda m \kappa}^{\rm outgoing} \otimes 
\bm{N}_{\lambda m \kappa}^{\rm regular}{}^* \right) \cr
&& + \sum_{\genfrac{}{}{0pt}{}{m=-\infty}{{m\neq 0}}}^\infty
\Gamma\left(|m|\right) \Gamma\left(|m|+1\right)
\bm{R}_{\lambda m \kappa}^{\rm outgoing} \otimes 
\bm{R}_{\lambda m \kappa}^{\rm regular}{}^*
\bigg\vert_{\lambda=\frac{1}{2i}}
\Bigg] \,,
\end{eqnarray}
where the regular (outgoing) function is evaluated at the point $\bm{r}_1$
or $\bm{r}_2$ with the smaller (larger) value of $|\theta|$.
Note that, as before, star indicates conjugation of the complex exponential
part of the function only.  Here the integral over $\displaystyle
\lambda =  \frac{1}{i}\left(\ell+\frac{1}{2}\right)$ represents the analytic
continuation of the sum over $\ell$.

For completeness, we also give the longitudinal mode
\begin{equation}
\bm{L}_{\lambda m \kappa} = -\frac{\sqrt{\lambda^2 +\frac{1}{4}}}{\kappa}
\nabla \left[
k_{i\lambda-\frac{1}{2}} (\kappa r) e^{im\phi} 
f_{\lambda m} (\bm{r}) \right]
\end{equation}
for this geometry.  If its contribution is added to the free Green's
function, the result is equal to the scalar Green's function 
$\displaystyle \frac{\kappa}{4 \pi} 
k_0\left(\kappa |\bm{r}_1 - \bm{r}_2|\right)$
times the identity matrix; without this contribution we obtain the
same scalar times the projection matrix onto the transverse components.

The full Green's function in the presence of the conducting boundary
$\mathfrak{G}$ is then given by the same expression with the
replacement
$\bm{\chi}^{\rm regular}{}^* \to 
\bm{\chi}^{\rm regular}{}^* + {\cal T}_{\lambda m}^\chi
\bm{\chi}^{\rm outgoing}{}^*$,
where $\bm{\chi} = \bm{M},\bm{N},\bm{R}$, again star indicates
conjugation of the complex exponential part of the function only, and 
${\cal T}_{\lambda m}^\chi$ is the corresponding $T$-matrix element
\cite{Maghrebi6867}
\begin{equation}
{\cal T}_{\lambda m}^N = 
-\frac{P^{-m}_{i\lambda -\frac{1}{2}}(\cos \theta_0)}
{P^{m}_{i\lambda -\frac{1}{2}}(-\cos \theta_0)}\,,
\qquad
{\cal T}_{\lambda m}^M =
-\frac{\frac{\partial}{\partial \theta_0}
P^{-m}_{i\lambda -\frac{1}{2}}(\cos \theta_0)}
{\frac{\partial}{\partial \theta_0}
P^{m}_{i\lambda -\frac{1}{2}}(-\cos \theta_0)} \,,
\qquad \hbox{and} \qquad
{\cal T}_{\lambda m}^R =
\frac{P^{-|m|}_{i\lambda -\frac{1}{2}}(\cos \theta_0)}
{P^{-|m|}_{i\lambda -\frac{1}{2}}(-\cos \theta_0)} \,,
\end{equation}
in terms of $\theta_0$, the half-opening angle of the cone.  Subtracting
the contribution from the free Green's function then cancels the term
with the regular solution, leaving only the product of outgoing
solutions in the interaction energy.

\section{Electromagnetic Cone Casimir-Polder energy}

After some algebra and simplification, we obtain the Casimir-Polder
interaction energy for an atom with isotropic and
frequency-independent polarizability $\alpha$  at distance $r$ from
the cone vertex and angle $\theta$ from the cone axis, with $|\theta| >
\theta_0$, as
\begin{eqnarray}
U(\bm{r}) 
&=& 2 \alpha \hbar c\int_0^\infty \kappa^2 \tr \left[
\mathfrak{G}(\bm{r},\bm{r},\kappa)-\mathfrak{G}_0(\bm{r},\bm{r},\kappa)\right]
d\kappa \cr
&=&-\frac{\alpha \hbar c}{2\pi} \int_0^\infty \kappa^3 \tr \Bigg[
\sum_{m=-\infty}^\infty 
\int_0^\infty \frac{\lambda \tanh \pi \lambda}
{\lambda^2 + \frac{1}{4}} d\lambda \left(
{\cal T}_{\lambda m}^M
\bm{M}_{\lambda m \kappa}^{\rm outgoing} \otimes 
\bm{M}_{\lambda m \kappa}^{\rm outgoing}{}^* - 
{\cal T}_{\lambda m}^N
\bm{N}_{\lambda m \kappa}^{\rm outgoing} \otimes 
\bm{N}_{\lambda m \kappa}^{\rm outgoing}{}^* \right) \cr
&& + \sum_{\genfrac{}{}{0pt}{}{m=-\infty}{{m\neq 0}}}^\infty
\Gamma\left(|m|\right) \Gamma\left(|m|+1\right)
{\cal T}_{\lambda m}^R 
\bm{R}_{\lambda m \kappa}^{\rm outgoing} \otimes 
\bm{R}_{\lambda m \kappa}^{\rm outgoing}{}^*
\bigg\vert_{\lambda=\frac{1}{2i}}
\Bigg] d\kappa 
\cr
&=& \frac{\alpha \hbar c}{2\pi r^2} \sum_{m=-\infty}^{\infty}
\int_0^\infty \kappa d\kappa \Bigg\{
\int_0^\infty d\lambda \lambda\tanh \pi \lambda
\Bigg[\Bigg(
\left(\frac{\partial}{\partial \theta}
P_{i\lambda -\frac{1}{2}}^m(-\cos \theta)\right)^2 
+ \frac{m^2}{\sin^2 \theta} P_{i\lambda -\frac{1}{2}}^m(-\cos
\theta)^2 \Bigg)\cr
&&\times \Bigg({\cal T}_{\lambda m}^N \Big(
\frac{\partial}{\partial r} 
\left(r k_{i\lambda -\frac{1}{2}}(\kappa r)\right)\Big)^2 -
{\cal T}_{\lambda m}^M \kappa^2 r^2
k_{i\lambda -\frac{1}{2}}(\kappa r)^2\Bigg)
\frac{1}{\left(\lambda^2 + \frac{1}{4}\right)}\cr 
&& + {\cal T}_{\lambda m}^N \left(\lambda^2 + \frac{1}{4}\right) 
P_{i\lambda -\frac{1}{2}}^m(-\cos \theta)^2 
k_{i\lambda -\frac{1}{2}}(\kappa r)^2\Bigg) \Bigg]
-\left(\frac{1+\cos \theta}{1-\cos\theta}
\right)^{|m|} \left(\frac{1-\cos \theta_0}{1+\cos\theta_0}
\right)^{|m|} \frac{2|m|}{\sin^2 \theta} e^{-2 \kappa r} \Bigg\} \cr
&=& \frac{\alpha \hbar c}{8r^4}\Bigg\{\sum_{m=-\infty}^{\infty}
\int_0^\infty d\lambda \lambda\sech \pi \lambda \tanh \pi \lambda \Bigg[
2 {\cal T}_{\lambda m}^N \left(\lambda^2 + \frac{1}{4}\right) 
P^{m}_{i\lambda -\frac{1}{2}}(-\cos \theta)^2 \cr
&& + \left({\cal T}_{\lambda m}^N - {\cal T}_{\lambda m}^M  \right)
\left(\left(\frac{\partial}{\partial \theta} 
P^{m}_{i\lambda -\frac{1}{2}}(-\cos \theta)\right)^2
+ \frac{m^2}{\sin^2 \theta} P^{m}_{i\lambda -\frac{1}{2}}(-\cos \theta)^2
\right)\Bigg] -\frac{1}{\pi} \frac{\sin^2 \theta_0}
{(\cos \theta - \cos\theta_0)^2} \Bigg\} \,,
\end{eqnarray}
where the last term arises from the ``ghost'' mode contribution.
Here we have used \cite{Gradshteyn}
\begin{equation}
\frac{\lambda^2 + \frac{1}{4}}{2r^2}
\int_0^\infty \kappa k_{i \lambda - \frac{1}{2}}(r\kappa)^2 d\kappa 
= \int_0^\infty \kappa^3 k_{i \lambda - \frac{1}{2}}(r\kappa)^2 d\kappa 
= \int_0^\infty \frac{\kappa}{r^2} \left[\frac{\partial}{\partial r}
\left(r k_{i \lambda - \frac{1}{2}}(r\kappa)\right)\right]^2 d\kappa 
= \frac{\pi}{4 r^4} \left(\lambda^2 + \frac{1}{4}\right) \sech \pi \lambda
\end{equation}
to carry out the integral over $\kappa$, with integrals involving
derivatives with respect to $r$ obtained by differentiating under the
integral sign.  The ghost term can be computed using the derivative of
a geometric series,
\begin{equation}
\sum_{m=1}^\infty
m \left(\frac{1+\cos \theta}{1-\cos\theta}\right)^{m} 
\left(\frac{1-\cos \theta_0}{1+\cos\theta_0}\right)^{m}
= \frac{\sin^2 \theta \sin^2 \theta_0}{4(\cos \theta -
\cos\theta_0)^2} \,,
\end{equation}
along with an elementary $\kappa$ integral.  

We can check this result numerically in the case of $\displaystyle
\theta_0 = \frac{\pi}{2}$, when it becomes the Casimir-Polder energy of
a particle at a distance $d = r |\cos \theta|$ from a conducting plane,
$\displaystyle U(\bm{r}) = -\frac{3 \alpha \hbar c}{8\pi d^4}$.  We
also note that we can simplify the difference of $T$-matrices to
\begin{equation}
{\cal T}_{\lambda m}^N - {\cal T}_{\lambda m}^M =
\frac{4\cosh \pi \lambda}{\pi \sin \theta_0}
\frac{1}{\frac{\partial}{\partial \theta_0} 
[P^{m}_{i\lambda -\frac{1}{2}}(-\cos \theta_0)^2]}
\end{equation}
by using the Wronskian relation between $P_\ell^m(z)$ and $P_\ell^m(-z)$.

Carefully taking the limit $\theta \to \pi$, we obtain the special
case where the particle lies on the cone axis.  Here the only
contributions arise from $m=-1,0,+1$, leading to a result that
simplifies to
\begin{eqnarray}
U(r) &=& \frac{\alpha \hbar c}{2\pi r^2} 
\int_0^\infty \kappa d\kappa \Bigg\{
\int_0^\infty d\lambda \lambda \left(\lambda^2 + \frac{1}{4}\right)
\tanh \pi \lambda \Bigg[
\left({\cal T}_{\lambda 0}^N - \kappa^2 r^2 {\cal T}_{\lambda 1}^M\right)
k_{i \lambda - \frac{1}{2}}(\kappa r)^2  \cr
&& + {\cal T}_{\lambda 1}^N
\left(\frac{\partial}{\partial r} \left(
r k_{i \lambda - \frac{1}{2}}(\kappa r)\right)\right)^2 \Bigg]
- e^{-2\kappa r} \tan^2 \frac{\theta_0}{2} \Bigg\} \cr
&=& \frac{\alpha \hbar c}{8 r^4} \left\{
\int_0^\infty d\lambda \lambda \left(\lambda^2 + \frac{1}{4}\right) 
\sech \pi \lambda \tanh \pi \lambda 
\left[
2 {\cal T}_{\lambda 0}^N + \left(\lambda^2 + \frac{1}{4}\right)
\left({\cal T}_{\lambda 1}^N - {\cal T}_{\lambda 1}^M\right)\right]
- \frac{1}{\pi} \tan^2 \frac{\theta_0}{2} \right\}
\end{eqnarray}
for the cone-particle interaction energy when $\theta = \pi$.  Here it
is helpful to obtain the result in the first line, before integration
over $\kappa$, because that result can straightforwardly be extended
to nonzero temperature and frequency-dependent polarization, as
will be described in more detail below.

Within this special case, it is illustrative to consider
$\displaystyle \theta_0 = \frac{\pi}{2}$, where the cone becomes a
plane, for which we have ${\cal T}_{\lambda 0}^N =-1$ and
$\displaystyle {\cal T}_{\lambda 1}^N = -{\cal T}_{\lambda 1}^M = 
-\frac{1}{\left(\lambda^2 + \frac{1}{4}\right)}$.  We can then use the
$\kappa$ integrals above along with the integrals
\cite{Gradshteyn,Oberhettinger_1961}\footnote{The second integral does
not appear to have been obtained previously.}
\begin{eqnarray}
\int_0^\infty k_{i \lambda - \frac{1}{2}}(\kappa r)^2 
\lambda \tanh \pi \lambda d\lambda &=& 
\frac{1}{2 \kappa r} e^{-2\kappa r} \cr
\int_0^\infty k_{i \lambda - \frac{1}{2}}(\kappa r)^2  
\lambda^3 \tanh \pi \lambda d\lambda &=& 
\frac{1}{2  \kappa r} \left(\kappa r + \frac{1}{4}\right) e^{-2\kappa r} \cr
\int_0^\infty \left( \frac{\partial}{\partial r}\left( r
k_{i \lambda - \frac{1}{2}}(\kappa r)\right) \right)^2  
\lambda \tanh \pi \lambda d\lambda &=& 
\frac{1}{2 \kappa r} \left(\kappa^2 r^2 - \kappa r + \frac{1}{2}\right) 
e^{-2\kappa r} \cr
\int_0^\infty d\lambda \lambda \left(\lambda^2 + \frac{1}{4}\right)
\sech \pi \lambda \tanh \pi \lambda &=& \frac{1}{2\pi} \,,
\end{eqnarray}
where again integrals involving derivatives with respect to $r$ are
obtained by differentiating under the integral sign, to do both
the $\kappa$ and $\lambda$ integrals explicitly and in either order and
obtain the standard results for the plane,
\begin{eqnarray}
U(r) &=& -\frac{\alpha \hbar c}{2\pi r^2} 
\int_0^\infty \kappa d\kappa \Bigg\{
\int_0^\infty d\lambda \lambda 
\tanh \pi \lambda \Bigg[
\left(\lambda^2 + \frac{1}{4} + \kappa^2 r^2\right)
k_{i \lambda - \frac{1}{2}}(\kappa r)^2  
+ \left(\frac{\partial}{\partial r} \left(
r k_{i \lambda - \frac{1}{2}}(\kappa r)\right)\right)^2 \Bigg]
+ e^{-2\kappa r} \Bigg\} \cr
&=& -\frac{\alpha \hbar c}{4\pi r^3} \int_0^\infty d\kappa 
\left(2\kappa^2 r^2+ 2 \kappa r + 1\right) e^{-2\kappa r} \cr
&=& -\frac{\alpha \hbar c}{2 r^4} \left[
\int_0^\infty d\lambda \lambda \left(\lambda^2 + \frac{1}{4}\right) 
\sech \pi \lambda \tanh \pi \lambda + \frac{1}{4\pi} \right] 
= -\frac{3\alpha \hbar c}{8 \pi r^4} \,,
\end{eqnarray}
where in the second line we have done the $\lambda$ integral first,
while in the third line we have done the $\kappa$ integral first.
The former expression shows that if the ghost contribution is grouped with
the electric modes, the contributions from the electric and magnetic
modes match the planar calculation individually as functions of
$\kappa$, and as a result reproduce the 5:1 ratio of their total
contributions \cite{Milton:19}.

\section{Anisotropic Polarizability}

By repeating the above calculation in the case where $\alpha$ is a
matrix, we can extend these results to the case of an anisotropic
particle.  We write the polarizability in the general form
\begin{equation}
\bm{\alpha} =
\begin{pmatrix}
\alpha_{\perp} \cos^2 \beta & \alpha_{xy} - i \gamma_z & \alpha_{xz} + i \gamma_y  \cr
\alpha_{xy} + i \gamma_z & \alpha_{\perp} \sin^2 \beta &  \alpha_{yz} - i \gamma_x \cr
\alpha_{xz} - i \gamma_y & \alpha_{yz} + i \gamma_x & \alpha_{zz}
\end{pmatrix} \,,
\end{equation}
where we have included both the symmetric and antisymmetric
(nonreciprocal) off-diagonal components.  Without loss of generality,
we take the particle to be at $\phi = 0$, so that it lies in the $xz$-plane.
In terms of these parameters, we obtain for the energy
\begin{eqnarray}
U(\bm{r})
&=& \frac{\hbar c}{8 r^4} \Bigg\{\sum_{m=-\infty}^{\infty}
\int_0^\infty d\lambda \lambda\sech \pi \lambda \tanh \pi \lambda \Bigg[
\cr &&
\left(\frac{\partial}{\partial \theta} 
P^{m}_{i\lambda -\frac{1}{2}}(-\cos \theta) \right)^2
\left((\alpha_{\perp} \cos^2 \beta \cot^2 \theta - 2 \alpha_{xz} \cot \theta + \alpha_{zz}) 
{\cal T}_{\lambda m}^N -\alpha_{\perp} \sin^2\beta \, {\cal T}_{\lambda m}^M\right)
\cr &&
+2 m \frac{\gamma_x + \gamma_z \cot \theta}{\sin \theta}
\left(\frac{\partial}{\partial \theta} 
P^{m}_{i\lambda -\frac{1}{2}}(-\cos \theta) \right)
P_{i\lambda -\frac{1}{2}}^m(-\cos \theta)
({\cal T}_{\lambda m}^N - {\cal T}_{\lambda m}^M)
\cr && +
P_{i\lambda -\frac{1}{2}}^m(-\cos \theta)^2
\Bigg( m^2 \left(2 \alpha_{xz} \cot \theta - \alpha_{\perp} \cos^2 \beta
\cot^2 \theta - \alpha_{zz} \right) {\cal T}_{\lambda m}^M
\cr && +
2\left(\frac{1}{4} + \lambda^2\right)
\left(\alpha_{zz}  \cos^2 \theta + 
\frac{2 m^2 \alpha_{\perp}}{(1 + 4 \lambda^2)} \csc^2 \theta \sin^2 \beta 
+ 2 \alpha_{xz} \cos \theta \sin \theta + \alpha_{\perp} 
\cos^2 \beta \sin^2 \theta \right) {\cal T}_{\lambda m}^N \Bigg)
\Bigg]
\cr && +
\frac{(2 \alpha_{xz} \cos \theta \sin \theta -\alpha_{\perp} +
(\alpha_{\perp} \cos^2 \beta -\alpha_{zz}) \sin^2 \theta) \sin^2 \theta_0}
{2\pi (\cos \theta - \cos \theta_0)^2}
\Bigg\}
\end{eqnarray}
which on the axis $\theta = \pi$ simplifies to
\begin{eqnarray}
U(\bm{r})
&=& \frac{\hbar c}{4 r^4} \Bigg\{
\int_0^\infty d\lambda \lambda \left(\frac{1}{4} + \lambda^2\right) 
\sech \pi \lambda \tanh \pi \lambda
\left[\alpha_{zz} {\cal T}_{\lambda 0}^N
+ \left(\frac{\alpha_{\perp}}{4} + \frac{\gamma_z}{2}\right)
\left(\frac{1}{4} + \lambda^2\right) (
{\cal T}_{\lambda 1}^N - {\cal T}_{\lambda 1}^M
)\right]
  -\frac{\alpha_{\perp} \tan^2 \frac{\theta_0}{2}}{4\pi}\Bigg\} \,. \cr &&
\end{eqnarray}
Of particular interest is the $\gamma_z$ term, which generates a
nonreciprocal torque around the $z$-axis.  Comparing the $\alpha_{zz}$
and $\alpha_{\perp}$ contributions also enables us to compare whether a
particle with a single polarization axis prefers to be aligned with or
perpendicular to the axis of the cone.

\section{Results and Discussion}

\begin{figure}[htbp]
\hfill
\includegraphics[width=0.475\linewidth]{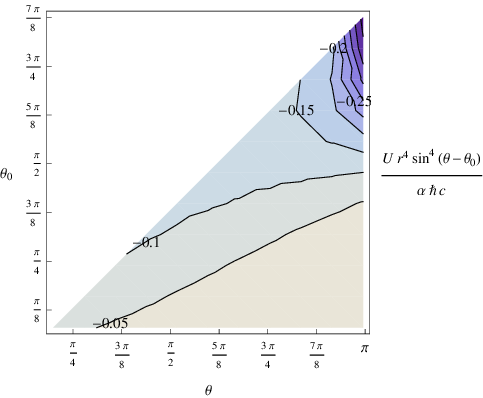} \hfill
\includegraphics[width=0.475\linewidth]{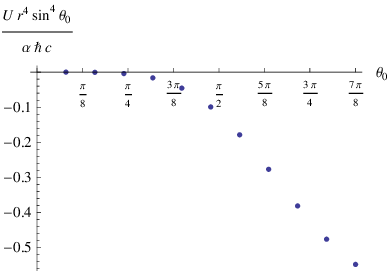} \hfill
\caption{Scaled Casimir-Polder interaction energy 
$\displaystyle \frac{U r^4 \sin^4 (\theta-\theta_0)}
{\alpha \hbar c}$ for an isotropic particle
as a function of $\theta$ and $\theta_0$ (left panel) and as a
function of $\theta_0$ for $\theta=\pi$ (right panel).}
\label{fig:CP}
\end{figure}

To visualize these results numerically, in Fig.~\ref{fig:CP} we plot
the Casimir-Polder interaction energy of an isotropic particle
scaled by the fourth power of $r
\sin(\theta-\theta_0)$, which gives the
perpendicular distance from the particle to the plane in the case
where $\displaystyle \theta-\theta_0 < \frac{\pi}{2}$.  For
$\displaystyle \theta_0 = \frac{\pi}{2}$, the result in units of
$\alpha \hbar c$ is $\displaystyle -\frac{3}{8\pi} \approx -0.1194$,
and past this inflection point as the the cone envelops the
particle, its interaction becomes much stronger.

We note that all of these calculations can be extended to nonzero
equilibrium temperature $T$, in which case the integral over $\kappa$
from $0$ to $\infty$
is replaced by $\displaystyle \frac{2 \pi k_B T}{\hbar c}$
times the sum over Matsubara frequencies 
$\displaystyle \kappa_n = \frac{2 \pi n k_B T}{\hbar c}$ for all
$n=0,1,2,3,\ldots$, where the $n=0$ contribution is counted with a
weight of $\displaystyle \frac{1}{2}$.  This term must be
considered carefully, since the Bessel function has a logarithmic
singularity as $\kappa \to 0$ for fixed $\lambda$.  For the special
case of $\displaystyle \theta_0 = \frac{\pi}{2}$, we can see
explicitly from the above that this singularity disappears when the
integral over $\lambda$ is done first, which should remain the case in
general.  In all of these calculations, one can also straightforwardly
move $\alpha$ inside the $\kappa$ integral or sum to model a
frequency-dependent polarizability.  However, introducing either or
both nonzero temperature and frequency-dependent polarizability
then requires the $\kappa$ sum or integral to be carried out numerically.

\section{Acknowledgments}
It is a pleasure to thank M.\ Kardar for suggesting this problem and,
along with K.\ Asheichyk, T.\ Emig, D.\ Gelbwaser, and M.\ Kr\"uger,
for helpful conversations and feedback. N.\ G.\ is supported in part by the
National Science Foundation (NSF) through grant PHY-2209582.

\bibliographystyle{apsrev}
\bibliography{casimirpolder}

\end{document}